\documentclass{WileyMSP-template}
\usepackage{amsmath}
\usepackage{amssymb}
\usepackage{siunitx}

\usepackage{booktabs}
\usepackage{dcolumn}

\usepackage{graphicx}

\usepackage{hyperref}

\newcommand{\mycomment}[1]{}
\begin{document}

\pagestyle{fancy}
\rhead{\includegraphics[width=2.5cm]{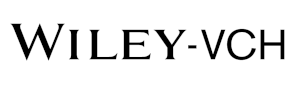}}

\title{Determining electron transfer flows between ligands and substrates with density functional theory in real time}

\maketitle

\author{Emmi P. Gareis}
\author{Stephan Kümmel*}

\begin{affiliations}
Theoretical Physics IV, University of Bayreuth, Germany\\
Email Address: stephan.kuemmel@uni-bayreuth.de
\end{affiliations}

\keywords{time-dependent density functional theory, electronic structure, charge transfer, first-principles simulations}

\begin{abstract}	
Charge transfer between a ligand and a substrate to which the ligand is anchored is a decisive process in many systems that convert light into other forms of energy. 
We discuss how the efficiency 
of such processes
can be quantified using density functional theory (DFT) in real time. 
Our approach focuses on the evaluation of the time-dependent density as the specific observable that can be exact in DFT. It avoids interpreting auxiliary quantities such as Kohn-Sham eigenvalues, can benefit from the computational efficiency of real-time propagation for large systems, and allows to
quantify the electronic connectivity on an atomistic scale.
We use it to study the electron transfer between phenylphosphonic acid and a WO$_3$ cluster, as the combination of WO$_3$ and anchored molecular complexes is promising for photoelectrochemical water splitting.
In line with common sense expectations we find that covalent anchoring of the ligand leads to a significantly higher flow of charge than
anchoring via a hydrogen bond. 
\end{abstract}

\section{Introduction}

Anchoring molecules to surfaces is an important aspect of several classes of systems that allow to convert sunlight into other forms of energy. A well known example are dye-sensitized solar cells \cite{holliman2018} that turn photons into electricity. An alternative option is to convert light into chemical energy, which is attractive because it can well be stored. 
Photoelectrochemical water splitting generating hydrogen as a clean energy carrier is one of the most promising processes of this type \cite{ahsaine2018}. Surfaces or nanoparticles of the semiconductor WO$_3$ are of great interest for photoelectrochemical water splitting because of WO$_3$'s attractive properties: a band gap in the visible range, a valence band maximum that is suitable for the oxygen evolution reaction, and chemical stability in acidic aqueous solutions \cite{costa2022current}. However, the oxygen evolution reaction on pristine WO$_3$ is hindered by large kinetic barriers. Attaching a molecular complex to the surface as a co-catalyst \cite{ben2022first,li2025,knodt2026determining} is one strategy for solving this problem -- if it can be achieved in such a way that charge can readily be transferred between the complex and the semiconductor. The latter condition is essential. Therefore, being able to determine 
whether such a charge transfer is likely to happen for a given combination of semiconductor and molecular ligand can put theory in the powerful position to help and guide experimental efforts for finding suitable materials.

In this paper we discuss how DFT in real time \cite{yabana1996,calvayrac2000,tddftbook2012,octopus2020,li2020,jakowski2025} can be used for obtaining insight into the efficiency of such charge-transport processes. The key idea is that a suitably chosen initial process triggers the flow of charge. Solving the (generalized) Kohn-Sham equations in real time one then monitors how much charge is being transferred between the molecular complex and the semiconductor as a function of time. Using real-time DFT for characterizing the quality of the electronic connection between a system and an attached ligand is conceptually attractive, because it avoids interpreting eigenvalues and orbitals of the Kohn-Sham system. The latter are introduced in DFT and its time-dependent extension as auxiliary quantities \cite{maitra2016}, whereas the decisive physical quantity is the electron density. Even with the exact exchange-correlation functional the Kohn-Sham orbitals and eigenvalues (except for the highest occupied one) will at best approximate physical quantities, whereas the density, on the other hand, will be exact. Therefore, extracting information about a system from analyzing its (time-dependent) density is the ideal way of using DFT. Real-time DFT has the further advantages of giving insight into the time scales on which processes happen, and it is a computationally attractive option for studying electron dynamics because it scales linearly with the number of electrons \cite{yabana1999}.

Our approach here is to be seen in the context of other real-time approaches that have been developed for studying conductivities or electron transport \cite{diventra2004,cheng2006,agosta2008,evans2009nanolett,zelovich2014,schaffhauser2016absBound,andermatt2018,he2018,cangi2023,oz2023}. In the present case, the initialization of the charge-transfer is less obvious than in the molecular transport regime, where the flow of charge is typically driven by an external bias. We here verify that different types of triggering the electron dynamics lead to very similar results. This demonstrates that one can indeed draw conclusions about the intrinsic electronic interaction between the ligand and the semiconductor from the real-time simulations. Furthermore, we deduce how one can estimate the initial, short-time conductivity that characterizes the connection between the ligand and the semiconductor.

While this paper is mainly intended to serve as a proof of concept, all our calculations are carried out for the practically relevant material combination of WO$_3$ as the substrate and phenylphosphonic acid as the molecular anchoring group. While finding ligands that anchor reliably on WO$_3$ is not a trivial task \cite{ben2022first,knodt2026determining}, phenylphosphonic acid has been identified as a promising candidate. Therefore, checking whether a flow of charge can be expected between these two systems, and for which types of anchoring, is a relevant question. 

Our paper is structured as follows. In section \ref{sec:TheoryAndBackground} we review the pertinent aspects of real-time DFT and explain the setup of our WO$_3$ and phenylphosphonic acid system. In section \ref{sec:Methods} we present as a first set of results three different methods for analyzing the ligand anchoring with real-time DFT. As a second set of results we present in section \ref{sec:ElectricalConductivity} methods for calculating a quantitative metric for characterizing the electronic connectivity of the anchoring. A summary and conclusions are offered in section \ref{sec:concl}.

\section{Real-time DFT and system setup} \label{sec:TheoryAndBackground}

\subsection{The basic concepts of density functional theory in real time} \label{sec:TDDFT}

In real-time DFT one solves the time-dependent Kohn-Sham equations
\begin{align}\label{eq:TD_KSeq}
  i \hbar \frac{\partial}{\partial t } \varphi_{j,\sigma} =
  \underbrace{
    \left(
        - \frac{\hbar ^ 2}{2 m} \nabla ^2 + v_\text{ion} +  v_\text{H} + v_{\mathrm{xc},\sigma}
  \right)
  }_{:= \hat{h}} \varphi_{j,\sigma}
\end{align}
by propagating the orbitals $\varphi_{j,\sigma}$ in real time \cite{yabana1996,calvayrac2000,octopus2020,li2020}. From the orbitals, which are strictly speaking just auxiliary quantities~\cite{runge1984density,tddftbook2012}, one obtains the time-dependent density
\begin{align}
	n(\boldsymbol{r},t) = \sum_{\sigma= \uparrow, \downarrow} \sum_{j=1}^{N_\sigma} \left| \varphi_{j, \sigma} (\boldsymbol{r},t ) \right|^2
\end{align}
as the decisive physical quantity, where $N_\sigma$ denotes the number of electrons with spin $\sigma$. 
In eq.\ (\ref{eq:TD_KSeq}) 
$v_\text{ion}$ denotes the interaction with the atomic cores. In the present work the orbitals $\varphi_{j,\sigma}$ refer to just the valence electrons, while the core electrons are implicitly contained in the pseudopotentials that also capture the atomic nuclei \cite{gareis2025submitted}. The Hartree ($v_\text{H}$) and exchange-correlation ($v_\text{xc}$) potential take into account the electron-electron interaction. The approximation chosen for $v_\text{xc}$ can play an important role for the electron dynamics, and certain systems can require a careful choice of $v_\text{xc}$ for an accurate description \cite{toher2005,ke2007,sicinteger}. For the present study that is intended mainly as a proof of the concept that ligand-semiconductor coupling can be quantified using real-time DFT, we mostly rely on the PBE~\cite{pbe} functional for reasons of computational efficiency. However, in addition we also use the LAK\cite{lak1,lak2} meta-GGA functional in some calculations. Meta-GGAs with a pronounced derivative discontinuity yield a more reliable description of band gaps and charge transfer \cite{TASK,eel2023,richter2023}. The meta-GGA calculations confirm that our results are robust with respect to this change of the exchange-correlation approximation. The technical details about our simulations with the real-space code BTDFT  \cite{SchelterKuemmelBTDFT} are reported in the Supplemental Material, which includes Refs. \cite{kresse1993ab,kresse1996efficiency,kresse1996comput,kresse1996software,kresse1994norm,kresse1999ultrasoft,troullier1991efficient,kleinman1980relativistic,bachelet1982relativistic,trepl2025time}.

\subsection{System setup to evaluate electronic transfer flows} \label{sec:TestSetup}

The anchoring group phenylphosphonic acid
(PPA) consists of a benzene ring attached to a phosphorus atom via a P–C bond, with the phosphorus center bearing three functional groups
-- two hydroxyls and one oxygen -- forming a tripod-like geometry. This structure enables two different types of surface interactions: On the one hand a weakly bound situation where a hydrogen atom from the anchoring group interacts with an oxygen atom on the surface, and on the other hand a more strongly bound situation with a covalent bond between a ligand oxygen and a surface tungsten atom.
Stable anchoring positions correspoding to these binding mechanisms were determined by geometry optimizations using ground-state density functional theory (DFT), cf.\ the Supplemental Material.

\begin{figure}[th]
	\centering
	\includegraphics[width=0.6\linewidth]{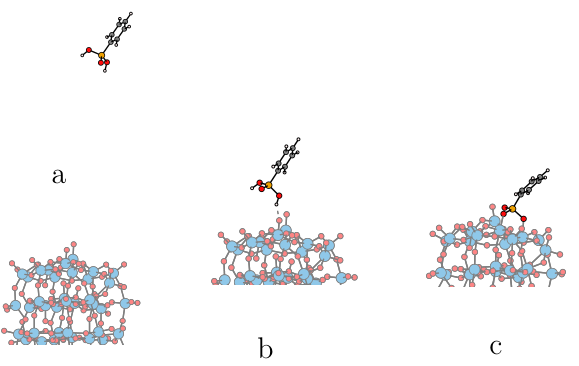}
	\caption{The three different systems which are studied in the following. (a): Large distance between the PPA and the cluster (b) PPA ligand attached to the cluster via a hydrogen bond (c) PPA ligand attached to the cluster via two covalent bonds. 
        }
	\label{fig:PPAon48WO3setup}
\end{figure} 
\textbf{Figure~\ref{fig:PPAon48WO3setup}} shows the configurations used in this study.
Panel (a) depicts a situation in which there is a large separation between the PPA and the WO$_3$-cluster. This serves as a reference and sanity-check: For this geometry, no charge transfer should be observed in any case. Thus, if a charge transfer is observed for this geometry with any of the methods that we discuss in the following sections, it must be considered a failure of the computational approach.
Panel (b) shows a situation in which the ligand is weakly bound to the cluster via a hydrogen bridge, and panel (c) shows a configuration in which it is bound covalently (with two bonds) to the cluster.
The common sense expectation is that (c) should allow for easier electron transfer between the ligand and the cluster than (b). Checking whether this can be corroborated and quantified on a first-principles basis with real-time DFT is the aim of the following sections.

\begin{figure*}[th]
	\includegraphics[width=\linewidth]{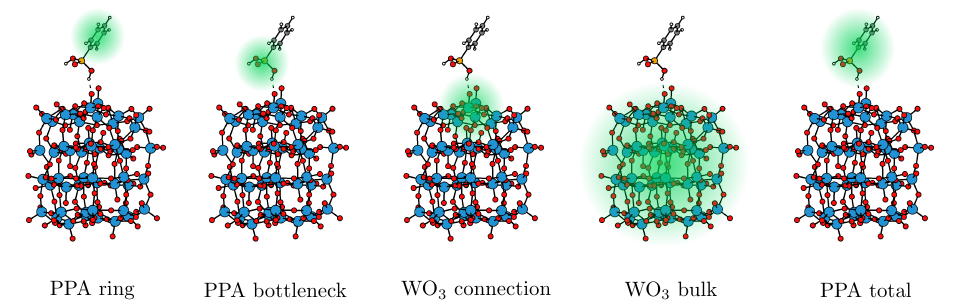}
	\caption{Schematic of the spatial partitioning used to analyze electron flow. The system is divided into five regions: (1) the aromatic ring of the PPA, (2) the tripod-like group around the phosphorus atom (referred to as the ``bottleneck''), (3) the surface connection region on the WO$_3$ cluster, and (4) the bulk of the cluster. Additionally, the sum of regions (1) and (2) defines the total region associated with the ligand (5). This partitioning enables the tracking of time-dependent electron number changes in each subsystem, facilitating a detailed analysis of charge transfer pathways and electron flow between the ligand and the cluster.}
	\label{fig:atomgroups}
\end{figure*}
Analyzing how the charge distribution between the ligand and the cluster changes as a function of time 
requires to define regions of space that are associated with the ligand and the cluster. For a more detailed analysis, it is helpful to additionally define intermediate regions through which the charge must flow when it moves between the ligand and the cluster.
\textbf{Figure~\ref{fig:atomgroups}} shows for the example of geometry (b) the five regions that we defined for this purpose: (1) the aromatic ring of the ligand, (2) the phosphorus-based tripod (referred to as the ``bottleneck''), (3) the surface connection region on the WO$_3$ cluster, (4) the bulk of the cluster, and (5) the region encompassing all of the ligand, enabling to monitor the total electron population within the ligand (see the Supplemental Material for the details of how these regions are defined).

For each region, the time-dependent electron number is computed by integrating the time-dependent electron density over the corresponding spatial volume:
\begin{align}\label{eq:Atomgroup}
	N_\text{g} =  \int_{V_{g}} n(\boldsymbol{r}, t) \, \, d^3 r 
\end{align}
with $g$ denoting one of the previously described regions, e.g., ``PPA total''.
Additionally, the difference density, which is defined as the difference between the time-dependent electron density and the ground-state density
\begin{align}\label{eq:DiffDens}
\delta n(\mathbf{r},t)=	n(\boldsymbol{r}, t) - n(\boldsymbol{r})_\text{\tiny groundstate},
\end{align} 
is analyzed to identify regions that become electron-depleted or electron-enriched during the charge transfer process.

While the idea of characterizing the 
strength of the electronic interaction between the ligand and the cluster
via the flow of charge 
with real-time DFT is straightforward, the question of how to set up such simulations requires thought. Our aim here is that the dynamics that one obtains from the real-time simulations should be characteristic for the intrinsic electronic structure and the coupling of the ligand-cluster-system. In other words, we want to be able to extract information about the coupling between ligand and cluster independently of how the flow of charge is initialized. To this end we discuss three different methods for initializing and simulating the charge transfer in the following. 

\section{Results 1: Methods for analyzing electron transfer flows} \label{sec:Methods}

\subsection{Initializing electron flow by a hole density} \label{sec:InstIon}

One scenario for photoelectrochemical water splitting is that light is absorbed in WO$_3$, creating electron-hole pairs. With the electrons removed via an electrode, the ``holes'' in the WO$_3$ provide the oxidation strength for the oxygen evolution reaction \cite{costa2022current}. A ligand attached to the WO$_3$-surface can make this process more efficient if the holes from the WO$_3$ can be transferred to the ligand to react with the water there. Independently of what exactly the interaction between the water and the ligand is going to be, a decisive step for this mechanism to work is that it must be possible to transfer a hole from WO$_3$ to the ligand through the anchoring group. In other words, it must be possible to transfer electrons from the ligand to WO$_3$.

We can probe this kind of electron transfer 
with real-time DFT by initially creating a ``hole'' in the WO$_3$ and then following the flow of charge in real time. 
This can be realized by removing one electron from the system, i.e., setting to zero the population of an orbital that was occupied in the initial $N_\mathrm{e}$-electron ground state, and then 
propagating the remaining orbitals of the initial $N_\mathrm{e}$-electron ground state using the $(N_e - 1)$-electron (generalized) Kohn-Sham Hamiltonian:
\begin{align}\label{eq:PropagatorInstIon}
	\varphi_{j} (t) = e^{-\frac{i}{\hbar}t \hat{h}_{(\text{N}_{e}-1)} } \, \varphi_{j, \text{ \tiny groundstate}}.
\end{align}

The orbital that is initially depopulated should be carefully chosen: If part of the probability density of the initially depopulated orbital would be on the ligand or in the anchoring region, the initialization would directly affect the dynamics that one wants to analyze. On the other hand, if one chooses an orbital that is confined to the WO$_3$-part of the system, then the electron flow through the anchoring group can evolve naturally in reaction to the charge situation in the WO$_3$, thus giving a better description of the situation that one wants to study. A further guideline for the choice of the orbital in our case comes in because we want to compare the dynamics in the three different ``anchoring'' situations (a), (b) and (c), cf.\ Figure \ref{fig:PPAon48WO3setup}, on the same footing. Therefore, the orbital that is initially depopulated should be as similar as possible in all three situations. This condition is in line with the condition that the orbital be confined to the WO$_3$ part of the system. 

\begin{figure}[th]
	\centering
	\includegraphics[width=0.28\linewidth]{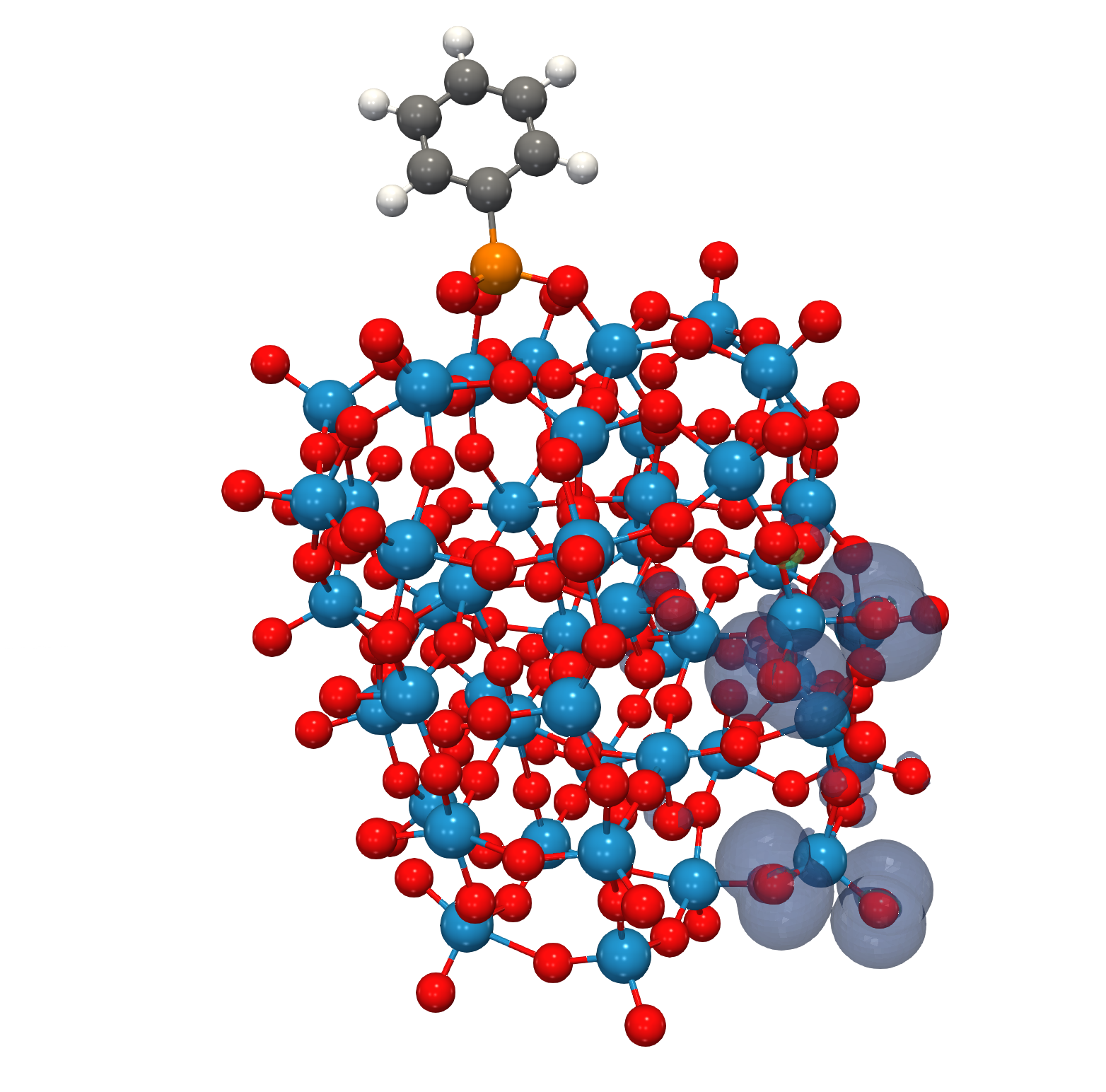}
	\includegraphics[width=0.25\linewidth]{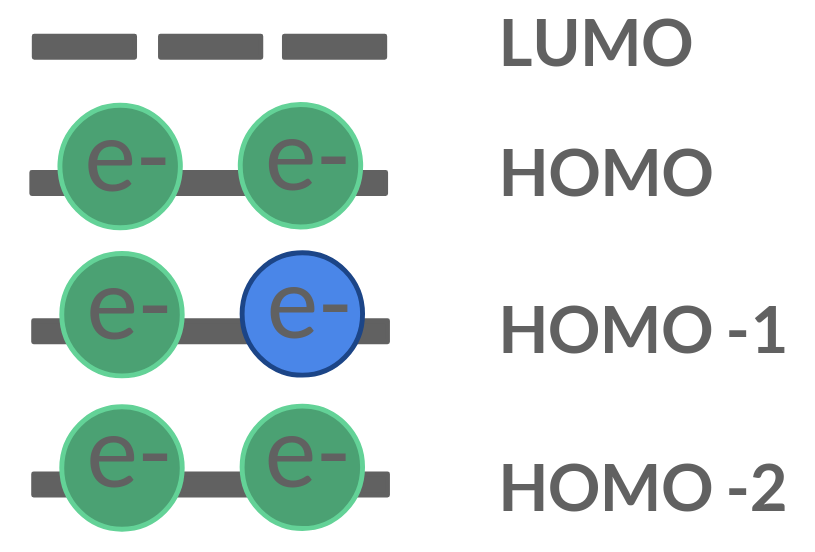}
	\caption{
	Schematic visualization of the orbital that was chosen to represent the ``hole''. The left part of the figure shows geometry (c) with O atoms in red, W atoms in blue, C atoms in grey, P atoms in orange, and H atoms in white. The lobes in the lower right part of the structure visualize the probability density of the HOMO-1 (isosurface 0.011 a$_0^{-3}$ of $|\varphi_\mathrm{HOMO-1}|^2$) of one spin channel, as illustrated in the right part of the figure. This orbital was depopulated to create the ``hole density''; see main text. 
	}
	\label{fig:selectedOrbital}
\end{figure}
For finding a suitable orbital we examined the highest occupied molecular orbitals of each system.  \textbf{Figure~\ref{fig:selectedOrbital}} depicts the situation for the geometry (c), i.e., the covalently-bound case. Here, the HOMO-1 fulfills the required conditions and was chosen for the initial depopulation representing the initial hole density.

\begin{figure}[th]
	\centering
	\includegraphics[width=0.8\linewidth]{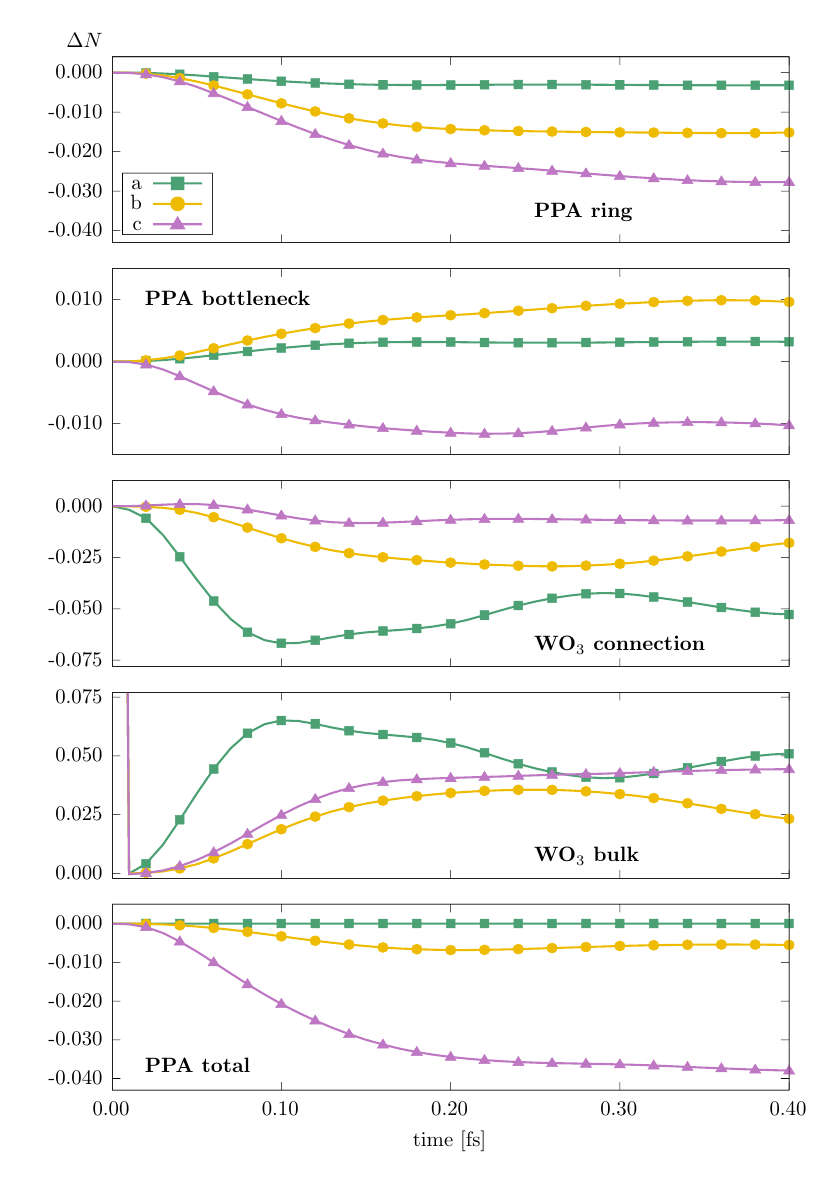}
	\caption{Integrated electron density as a function of time after the hole density initialization for the different systems shown in Figure \ref{fig:PPAon48WO3setup} (green squares: (a); yellow circles: (b); violet triangles: (c) ) and for the different regions of space defined in Figure \ref{fig:atomgroups}. From top to bottom: PPA ring, PPA bottleneck, WO$_3$ connection, and WO$_3$ bulk. The lowest panel shows the total PPA region.
	}
	\label{fig:instIon}
\end{figure}

The five different panels in \textbf{Figure~\ref{fig:instIon}} show the evolution of the integrated electron density in the five regions that were defined in Figure \ref{fig:atomgroups}. Let us start by first looking at the fourth plot from the top in Figure~\ref{fig:instIon} labeled ``WO$_3$ bulk''. The three types of colored curves and symbols correspond to the three different geometries described in Section~\ref{sec:TestSetup} and illustrated in Figure~\ref{fig:PPAon48WO3setup}.  The depicted dynamics is triggered by the removal of an electron from the selected orbital in the first time step. This initial electron removal is evident as a sharp drop immediately after $t = 0$ (only one color visible as the three curves are initially right on top of each other). Subsequently, the electron density reacts to this creation of a ``hole''. Changing to the lowest panel in Figure~\ref{fig:instIon} that shows the density change in the ligand (label ``PPA total'') and looking at the green squares that mark the results for situation (a) in which PPA and WO$_3$ are far apart, we see that the charge on the ligand stays constant. Thus, the sanity check reassuringly gives the correct result that no charge flows from the ligand to the cluster when they are far apart. Looking at the green lines in the upper two panels (labeled ``PPA ring'' and ``PPA bottleneck'') we see that there is a small redistribution of electron density from the PPA ring to the PPA bottleneck. This is explained by the Coulomb interaction: After an electron has been removed from the cluster (the initial ``creation of a hole''), the electron charge in the ligand is being slightly polarized towards the cluster. The substantial charge dynamics within the cluster that is seen in the panels labeled  ``WO$_3$ bulk'' and ``WO$_3$ connection'' is also intuitively plausible: After the creation of the hole in the lower part of the cluster, cf.\ Figure \ref{fig:selectedOrbital}, some electron density from the top part (i.e., the ``connection'' region) moves to compensate for the hole. This leads to a density osciallation within the cluster with a net effect of shifting some charge from the ``connection'' region to the ``bulk'' region. 
\clearpage

Next, let us analyze the weakly bound situation (b), represented by the yellow circles. In the lowest panel ``PPA total'' we see that a little bit of charge now leaves the PPA ligand. The effect, however, is relatively small. The upper two panels reveal that the charge moves from the ``PPA ring'' region to the ``PPA bottleneck'' region. This corresponds to a noticeable polarization of the ligand, and some flow of charge into the WO$_3$ cluster. Within the WO$_3$ we see the same effect as before: On average, charge is redistributed from the ``connection'' region to the ``bulk'' region as the density redistributes within the cluster to compensate the hole. 

The most interesting scenario (c) in which the ligand is covalently bound to the WO$_3$ is shown by the violet triangles. The lowest panel shows that now a substantial amount of charge leaves the PPA ligand. The upper two panels further show that both the ``ring'' and the ``bottleneck'' region lose charge. As a consequence, the WO$_3$ ``connection'' region is far less depleted of charge than in the previous cases, and the charge that left the PPA ends up in the ``WO$_3$ bulk'' region. Thus, these simulations confirm that the covalent anchoring (c) of PPA on WO$_3$ can allow for a transfer of charge.

\begin{figure*}[h]
	\includegraphics[width=\linewidth]{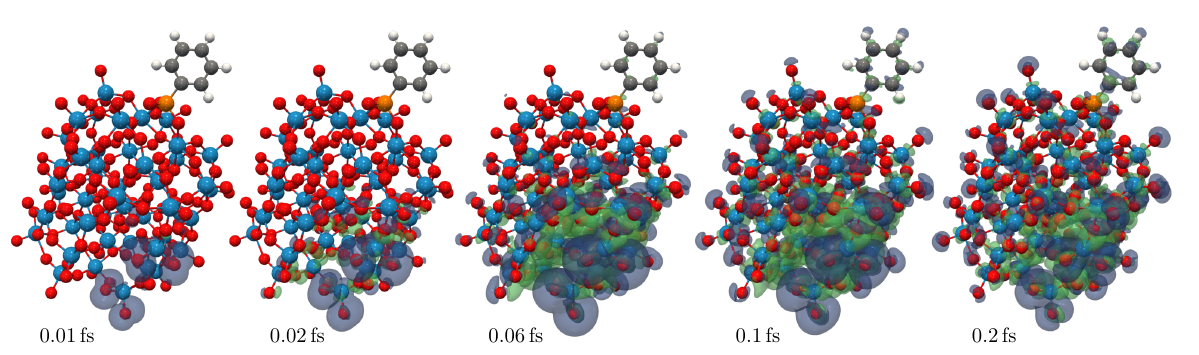}
	\caption{The difference density for geometry (c) after the hole initialization as a function of time. Blue regions indicate electron depletion, green regions electron accumulation (isosurface $\pm 0.011$ a$_0^{-3}$). The initially depleted region is filled by electrons from the ligand, as seen in the growing green region at the ligand-cluster connection, and the growing blue region on the ligand ring.}
	\label{fig:diffDens_instIon}
\end{figure*}
The conclusion that charge is being transferred from the ligand to the cluster in situation (c) is also confirmed by the difference density, which is visualized as a function of time in \textbf{Figure~\ref{fig:diffDens_instIon}}. Blue regions indicate electron depletion and green regions denote electron accumulation relative to the ground state. Initially, the depopulated orbital is highly prominent, manifesting as an extended blue region in the lower part of the cluster. As time progresses, electron density flows back into this region from across the system, including the ligand. The involvement of the ligand is evident from the growing green region that develops right where the ligand connects to the cluster, and the blue region developing on the ligand ring. 

It is reasonable to check how sensitive these results are to the employed exchange-correlation approximation.
Therefore, we repeated the calculations with the meta-GGA LAK~\cite{lak1,lak2} which was derived from first-principles and exact constraints, and yields much more realistic band gaps than PBE, and an improved description of charge transfer. The results reassuringly confirm the PBE claculations, as discussed in detail in the Supplemental Material.

One might wonder in how far the conclusions about the ability to transfer charge through a given molcular anchoring configuration are influenced by the specific way in which the flow of charge is triggered in the  real-time DFT simulations. Therefore, in the following we initiate the charge transfer in two other, completely different ways.

\subsection{Initializing electron flow by an electrical field} \label{sec:E-Field}

An alternative approach to initiate electron transfer from the ligand to the cluster core -- and thus determine in how far a certain anchoring type is suitable for electron transfer -- is the application of an external, homogeneous, and constant electric field. This approach does not model the ``hole'' interpretation that is usually invoked in the context of the oxygen evolution reaction on WO$_3$ \cite{costa2022current}. However, in the present context this is a great advantage: Using a completely different way of initializing the electron dynamics will allow to check whether the conclusions about the coupling between the ligand and the cluster are independent of the initialization. The electrical field approach is attractive also from another perspective: Future work will aim to use the real-time approach for ligands that are attached to extended crystal surfaces, not finite clusters. Such calculations will need to invoke periodic boundary conditions. Then, charging the semiconductor as done for the cluster in Section \ref{sec:InstIon} will not be possible. The electrical field approach, however, promises to be useful also in the case of periodic boundary conditions.

We want the electrical field to be able to drive charge from the ligand to the cluster, i.e., it should be perpendicular to the cluster surface. As the surface of the cluster is not ideally flat, we determined the perpendicular direction via the plane that is defined by the third layer of $W$ atoms. This layer is quite flat, as one can see in Figure \ref{fig:atomgroups}. We defined this plane to be the $x$-$y$-plane, and for defining it we averaged over the slightly different $z$-coordinates of the $W$ atoms defining the plane. The electrical field was then chosen to be in the $z$-direction, pointing from the cluster to the ligand.
Starting from the ground-state orbitals we switched on a homogenous electrical field of strength $E_\mathrm{ext}$ (see the Supplemental Material for details)  and propagated the orbitals according to
\begin{align}\label{eq:PropagatorElectricField}
	\varphi_{i} (t) = e^{-\frac{i}{\hbar}t (\hat{h} -q E_\text{ext} z ) } \, \varphi_{i, \text{ \tiny groundstate}}.
\end{align}
If the ligand and the cluster are sufficiently connected, the field can drive electrons from the ligand to the cluster, potentially inducing a current across the ligand-cluster interface.

\begin{figure}[th]
	\centering
	\includegraphics[width=0.8\linewidth]{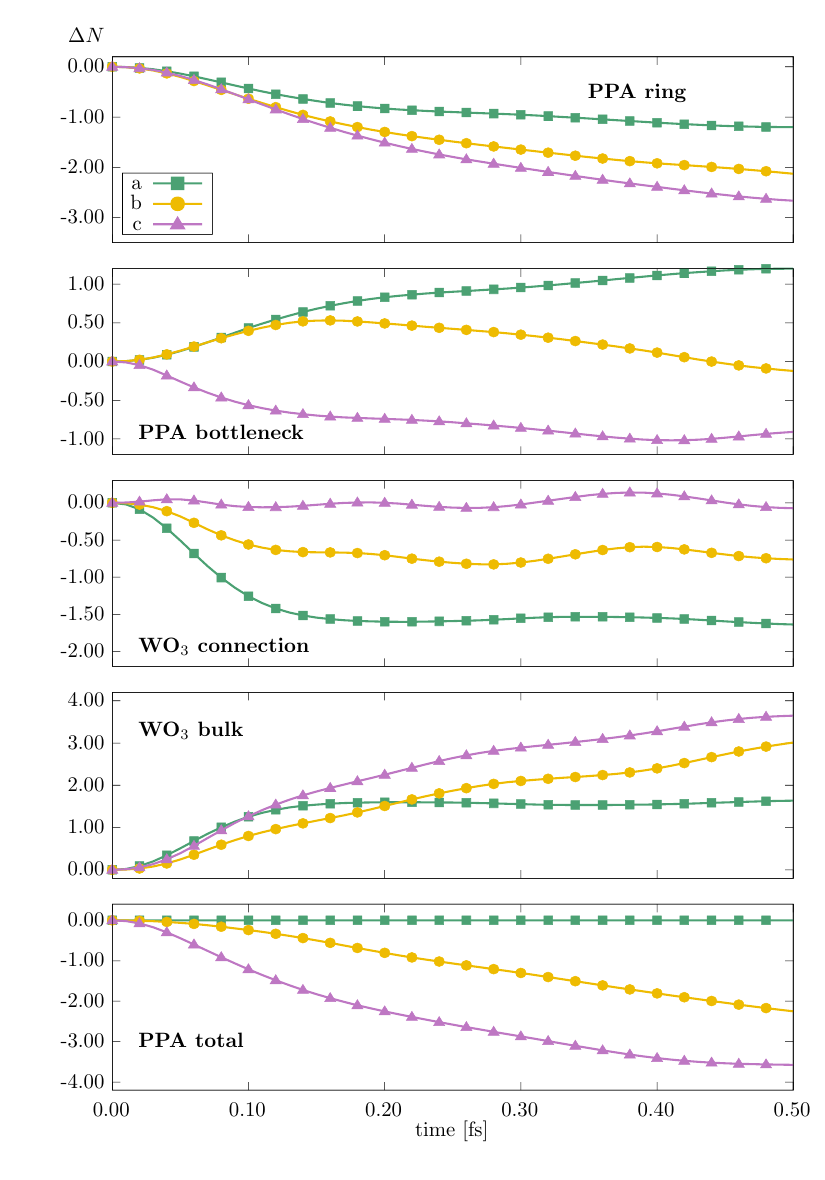}
	\caption{
	Integrated electron density as a function of time under the influence of an external electrical field for the different systems shown in Figure \ref{fig:PPAon48WO3setup} and for the different regions of space defined in Figure \ref{fig:atomgroups}. The labels and color codes are the same as in Figure \ref{fig:instIon}.
	}
	\label{fig:E-field}
\end{figure}

\textbf{Figure~\ref{fig:E-field}} shows the evolution of the integrated electron density in analogy to Figure~\ref{fig:instIon}, but here for the case that the dynamics is triggered by the external electrical field.
Qualitatively, one observes in the lowest panel ``PPA total'' similar effects as in Section \ref{sec:InstIon}: The green squares show that again no charge is being transferred from the ligand to the cluster when the separation between them is large, and considerably less charge is being transferred in the weakly bound situation (b) (yellow) than in the strongly bound situation (c) (violet). However, the scale on the vertical axis is very different. With the electrical field of the chosen strength, much more charge is being transferred than in the case of the ``hole'' initialization. In the well-connected situation (c), this charge accumulates inside of the WO$_3$ cluster, cf.\ panel ``WO$_3$ bulk'', and is missing in both parts of the PPA ligand (as seen in the upper two panels). The polarization that the field induces builds up an internal electrical field that counters the externally applied field. The internal field slows down and eventually stops the further transfer of charge from the ligand to the cluster. Consequently, the slope of the violet curve in the lowest panel ``PPA total'' decreases for larger times. The information about the coupling between the ligand and the cluster is therefore most clearly encoded before this saturation point is reached, i.e., around ca.\ $\SI{0.1}{\femto \second}$. This regime is also the one that we will focus on when assessing a measure for the electrical conductivity of the ligand-cluser connection in Section~\ref{sec:ConductivityE-Field}.

\clearpage

\begin{figure*}[h]
	\includegraphics[width=\linewidth]{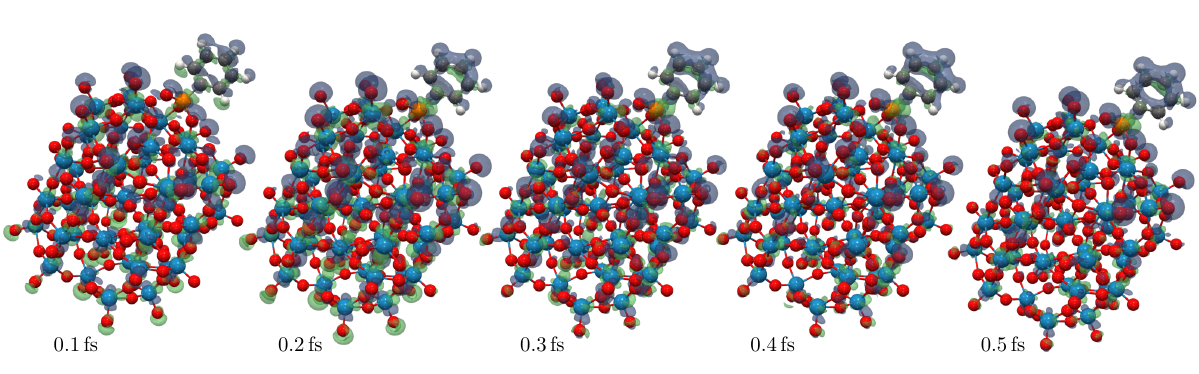}
	\caption{Time evolution of the difference density under an external electric field . Blue denotes electron depletion, green electron enrichment. Shown are the isosurfaces of $\pm 0.33$ a$_0^{-3}$.
        }
	\label{fig:diffDens_E-Field}
\end{figure*}
\textbf{Figure~\ref{fig:diffDens_E-Field}} shows the corresponding time evolution of the difference density for the strongly bound configuration (c). Blue regions indicate electron depletion and green regions indicate electron enrichment relative to the ground state.
Within the first ca.\ $\SI{0.2}{\femto\second}$, the number of electrons localized at the bottom of the cluster significantly increases, as evidenced by the green regions. In the further course of time, a substantial amount of charge transfers from the ligand to the cluster, as demonstrated by the extensive blue regions spreading across the benzene ring and the tripod-like connection. Within the cluster, the additional electron density then on average distributes relatively uniformly. One observes a slight back-and-forth flow of charge within the cluster. This is understandable as the external fiel triggers quantum mechanical dynamics, and this involves scattering and interference.
The visualization of the difference density also confirms the previously discussed slowing down of the charge transfer that is a result of the polarization buildup.
In the following, we thus discuss a third method for driving the charge dynamics that avoids the accumulation of electron density in the lower region of the cluster.

\subsection{Initializing electron flow by an absorbing boundary} \label{sec:AbsBound}

An established method for reducing the effects of charge accumulation and backscattering is the use of absorbing boundary conditions. An absorbing boundary can simulate an open quantum system in a finite-system setup by numerically realizing an electron drain. In the present context, an absorbing boundary that extracts electrons from the lower part of the cluster can be used to mimic the attraction that ``holes'' in the WO$_3$ or a positive electrode have on the electrons in the ligand. As stated before, we employ this setup as a third test for whether information about the electronic connection and the transfer of charge between cluster and ligand can be extracted reliably from real-time DFT.

An absorbing boundary can be introduced via a complex-valued potential that is added to the Hamiltonian featuring in the propagator, i.e.,
\begin{align}\label{eq:PropagatorAbsBound}
	\varphi_{i} (t) = e^{-\frac{i}{\hbar}t (\hat{h} - i \tilde{v}) } \, \varphi_{i, \text{ \tiny groundstate}}.
\end{align}
For the absorbing potential we here use a form that has been suggested in Ref.~\cite{manolopoulos2002derivation},
\begin{align}\label{eq:AbsBoundary_Mano}
	\tilde{v}(\boldsymbol{r}) =	k \left[ \left(\frac{\Delta x}{x_2 -2x_1 + \boldsymbol{n}\cdot \boldsymbol{r}}\right)^2 +\left(\frac{\Delta x}{x_2 - \boldsymbol{n}\cdot \boldsymbol{r}}\right)^2  -2 \right],
\end{align}
and which has been successfully used in previous work~\cite{manolopoulos2002derivation,dauth2016manoBTDFT,schaffhauser2016absBound}. It ensures efficient absorption while minimizing reflection at the boundary. The parameters in Equation (\ref{eq:AbsBoundary_Mano}) are
$k = (4 \hbar^2)/(2mc^2)[(2\pi)/(\Delta x)]^2$ with $c= 2.62$, and $x_1 = \boldsymbol{n} \cdot \boldsymbol{r_1}$ defines the plane where the absorbing layer begins and $x_2 = \boldsymbol{n} \cdot \boldsymbol{r_2}$ the one where it ends. The vector $\boldsymbol{n}$ denotes the normal vector of the absorbing layer and therefore specifies the spatial orientation of the absorbing boundary. Consequently, $\Delta x = x_2 -x_1$ defines the thickness of the absorbing layer.

\begin{figure}[th]
	\centering
	\includegraphics[width=0.8\linewidth]{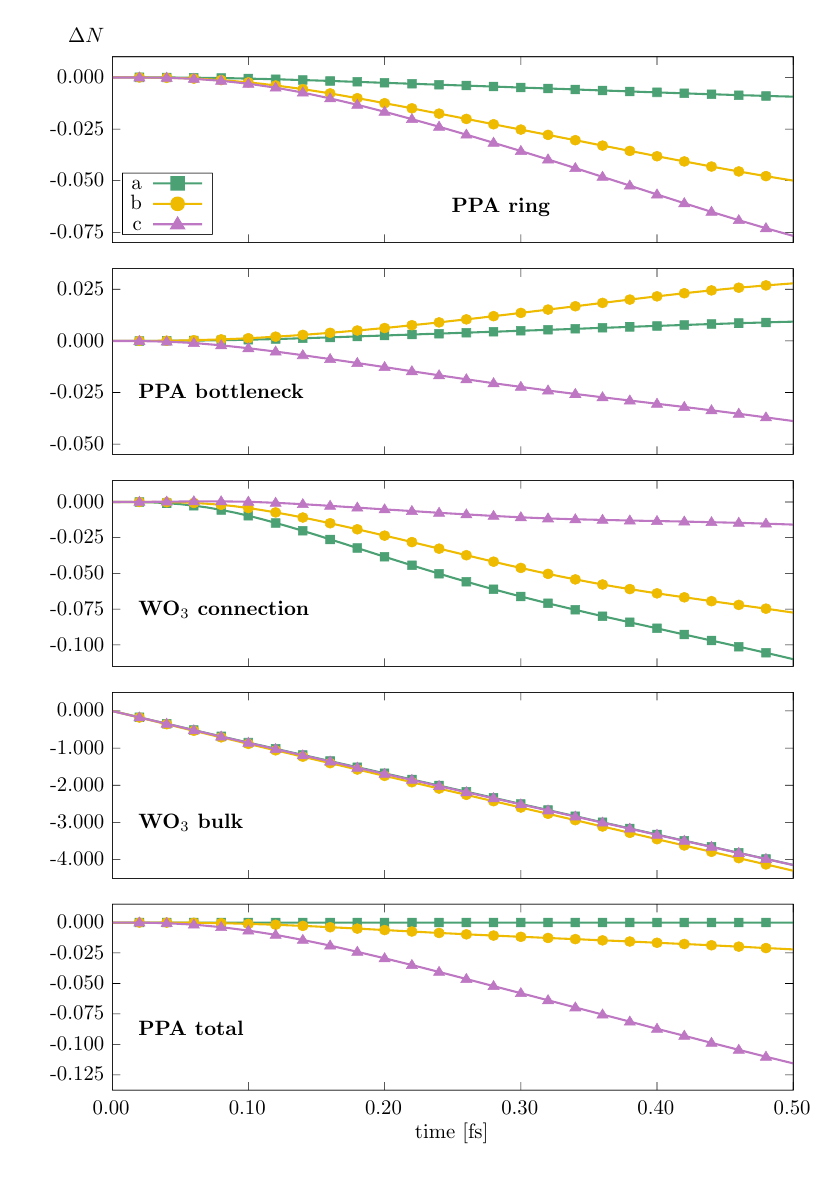}
	\caption{
	Integrated electron density as a function of time under the influence of an absorbing boundary for the different systems shown in Figure \ref{fig:PPAon48WO3setup} and for the different regions of space defined in Figure \ref{fig:atomgroups}. The labels and color codes are the same as in Figure \ref{fig:instIon}.
        }
	\label{fig:AbsBound}
\end{figure}
\clearpage

We chose the boundary layer such that electron removal occurs specifically from the lower region of the cluster, thereby initiating electron transfer from the ligand to the cluster.
Via the parameters of the absorbing potential one can control the rate of electron extraction. The detailed parameters that we used are reported in the Supplemental Material. 

\textbf{Figure~\ref{fig:AbsBound}} shows the evolution of the electron density under the influence of the absorbing boundary. The density is integrated over the same regions, for the same geometries, and plotted in the same way as in sections \ref{sec:InstIon} and \ref{sec:E-Field}. The boundary was set up such that it continuously removes electron density from the region ``WO$_3$ bulk'', as seen in the fourth panel from the top, such that a total of four charges is removed over the entire simulation time. This might seem a lot on first sight. However, looking at the upper two panels ``PPA ring'' and ``PPA bottleneck'', and the lowest panel ``PPA total'', shows that the density changes induced in these regions are still quite small. Thus, the seemingly strong driving via the absorbing boundary at the lower end of the cluster is useful to obtain meaningful and numerically stable signals in the other regions. We made sure that the removal of charge was done consistently for the three configurations (a), (b), and (c), as confirmed by the well-aligned curves in the panel ``WO$_3$ bulk''. In this way we ensure that the simulation results for the three geometries (a), (b), and (c) can be quantitatively compared.

Looking at the lowest panel ``PPA total'' shows that qualitatively, the conclusions about the coupling between ligand and cluster are consistent with the conclusions drawn in the previous sections: No electron transfer is observed for the large separation (a), and the transfer in the covalently bound system (c) is considerably larger than in the weakly bound system (b).  

Looking at the yellow lines in the upper two panels ``PPA ring'' and ``PPA bottleneck'' shows that there is a significant flow of charge from the benzene ring, but a substantial fraction of this charge accumulates in the ``bottleneck'' region. This clearly demonstrates that the weak binding is not promising for electronically connecting ligands to a substrate. This finding is again in agreement with the conclusions from the previous sections.

However, there are also clear differences between the curves obtained with the absorbing boundary method and the curves in Figures \ref{fig:instIon} and \ref{fig:E-field}: The absorbing boundary leads to significantly smoother dynamics, i.e., the absorbing boundary comes closest to realizing a ``steady-state''-like flow of charge during the studied simulation time.

\begin{figure*}[h]
	\includegraphics[width=\linewidth]{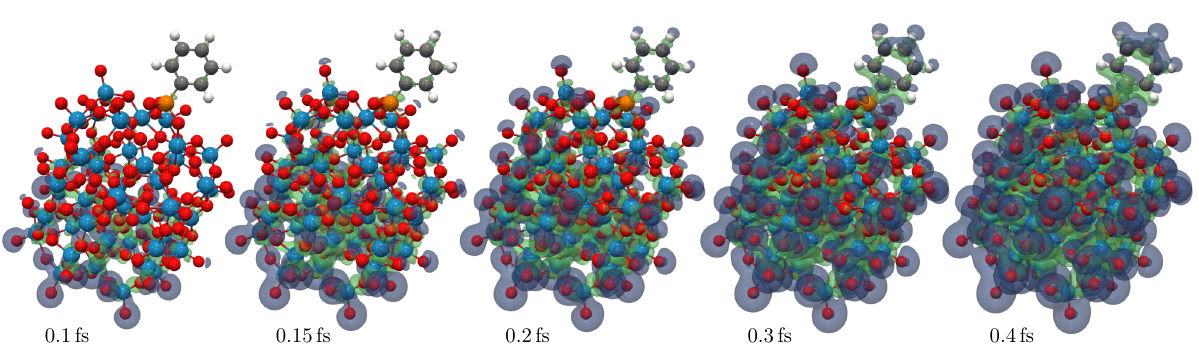}
	\caption{
Time evolution of the difference density under an absorbing boundary for configuration (c). Blue denotes electron depletion, green electron enrichment. Shown are the isosurfaces of $\pm 0.011$ a$_0^{-3}$.
        }
	\label{fig:diffDens_absBound}
\end{figure*}
\textbf{Figure~\ref{fig:diffDens_absBound}} displays the corresponding time evolution of the difference density for the strongly bound configuration (c). As previously, blue regions indicate electron depletion and green regions denote electron enrichment relative to the ground state. The absorbing boundary continuously removes density from the lower region of the cluster, and this leads to the steadily growing blue regions in this area. The electron depletion propagates relatively uniformly across the cluster, spreading from the bulk towards the ligand. At $\SI{0.3}{\femto\second}$, a green region at the phosphorus-based tripod indicates electron enrichment, while blue regions on the benzene ring highlight electron depletion -- clearly signaling that electrons flow from the ligand’s ring into the cluster via the bottleneck. Once again, the observed charge redistribution provides unambiguous evidence of the electron exchange between the cluster and the ligand, in agreement with the previous findings.

Up to now, the conclusions about the electronic connectivity between ligand and cluster were somewhat qualitative, as the details of the observed dynamics were influenced by the way in which the dynamics was initialized. 
In the following we discuss how one can calculate 
a quantitative metric for the electronic conductivity that characterizes the connection between the ligand and the cluster in different bonding situations.

\section{Results 2: Quantitative measures of the anchoring connectivity}\label{sec:ElectricalConductivity}

\subsection{Analysis of the anchoring with the constant field method} \label{sec:ConductivityE-Field}

We here use the term ``conductivity'' in the context of characterizing the electronic connection between the ligand and the cluster in the different bonding situations. This term is fitting, as 
our quantity $\sigma$, see below, has the unit of a conductivity. We would like to stress, though, that this $\sigma$ is not the same as the conductivity that one typically is thinking about in solid-state physics. The latter is governed by effects such as the electron-phonon interaction and lattice defects, whereas our present setup focuses solely on quantifying the electronic structure connection between the ligand and the cluster.
In the following we discuss how this type of ``conductivity'', which one might alternatively call ``connectivity'', can be calculated for all three of the previously introduced TDDFT methods of initializing a flow of charge through an anchoring group.

The derivation for the constant external electrical field approach is the most intuitive of the three, and therefore we begin with this case. As previously discussed in section \ref{sec:E-Field}, charge flows from the ligand into the cluster under the influence of an external electrical field. The outflow of charge from the ligand 
corresponds to the current of interest.
We define an electrical conductivity $\sigma$ that characterizes the connection between ligand and cluster as the proportionality constant between the current density $\boldsymbol{j}$,
defined as the current $I_\mathrm{lig}$ per area $A$,
and the externally applied electrical field $\boldsymbol{E}_\text{ext}$ driving the current, i.e.,
\begin{align}
\label{eq:j}
	\boldsymbol{j} = j \, \boldsymbol{e}_z = \frac{I_\text{lig}}{A} \, \boldsymbol{e}_z =  \sigma \, E_\text{ext} \boldsymbol{e}_z. 
\end{align}
Since the external field is along the $z$-axis, the equations reduce to the z-components. 
$I_\mathrm{lig}$ can be calculated straightforwardly from the change of the charge $ Q_\text{lig}$ of the ligand: Integrating the density over the volume that is associated with the ligand, i.e., the region ``PPA total'' (see the Supplemental Material for the details of the definition of the volume) yields $N_\text{lig}$, and thus
\begin{align}\label{eq:jSigmaE_E-Field}
	 I_\text{lig} = \frac{ d | N_\text{lig}| e }{d t}=\frac{ d Q_\text{lig} }{d t} = \sigma A E_\text{ext}.
\end{align}
In equation (\ref{eq:jSigmaE_E-Field}) $e$ is the elementary charge, and equation (\ref{eq:j}) has been used in the last step on the right-hand side. $A$ denotes the cross-sectional area of the connection between the ligand and the cluster. On the atomistic scale that we are looking at, such an area can hardly be defined unambiguously. This, however, is not a problem, as $\sigma$ and $A$ together define the quality of the electronic connection between the ligand and the cluster. The product
\begin{align}\label{eq:SigmaA}
	  \sigma \cdot A =  \frac{I_\text{lig}}{E_\text{ext}}.
\end{align}
can thus be used to characterize the ligand-cluster connectivity. The decisive observation is that this product can straightforwardly be obtained from the real-time simulation, as both quantities on the right-hand side of equation (\ref{eq:SigmaA}) are known: $E_\text{ext}$ is a parameter of the simulation, and $I_\text{lig}$ follows from the left-hand side of equation (\ref{eq:jSigmaE_E-Field}). 

In a time-dependent simulation, the right-hand side of equation (\ref{eq:SigmaA}) will not be a constant: When the field is switched on, there will be initial effects and the current will increase as a function of time. Then, as time progresses, the counter-field that we discussed in section \ref{sec:E-Field} will build up, and it will start to slow down the further transfer of charge. Consequently, the best point in time for evaluating equation (\ref{eq:SigmaA}) is shortly before the slow-down of the charge transfer fully sets in.

\begin{figure}[th]
	\centering
	\includegraphics[width=0.7\linewidth]{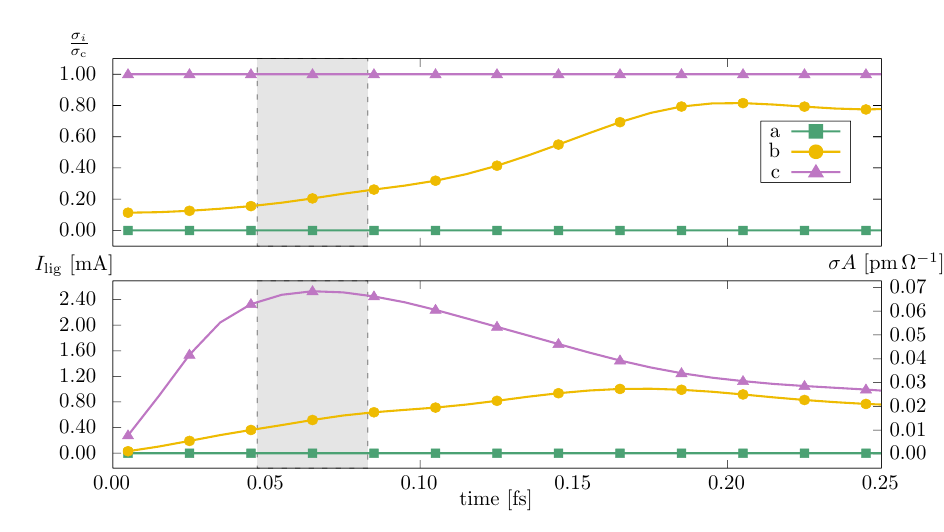}
	\caption{Bottom: Time evolution of $I_\text{lig}$ (left vertical axis) and of $\sigma A$ (right vertical axis) under an external electrical field. These two quantities have the same time dependence and differ only in magnitude and unit, cf.\ Equation (\ref{eq:SigmaA}). Top: Relative conductivity ratios. 
	The labels and color codes are the same as in Figure \ref{fig:instIon}. The hatched area indicates the range of time in which the effects of the counter field become prominent. See main text for discussion.
	}
	\label{fig:IandSigma_E-Field_0.15}
\end{figure}
The bottom panel of \textbf{Figure~\ref{fig:IandSigma_E-Field_0.15}} shows $I_\mathrm{lig}$ from Equation (\ref{eq:jSigmaE_E-Field}) as a function of time. For the geometry (a) with the large separation, one does not observe any current; the green squares are at zero for all times. For the covalently bound situation (c) that is represented by the violet triangles one observes an initial sharp rise of the current. Then, the slope 
becomes smaller before 0.1 fs are reached. The curve is such that one can hardly identify one exact point in time when the current starts to level off. Therefore, at the present stage of the analysis there is a certain ambiguity in determining the point in time at which the current best characterizes the intrinsic properties of the ligand-cluster connection. One could argue that any of the times within the hatched region in Figure~\ref{fig:IandSigma_E-Field_0.15} could be chosen. The reassuring observation is that for all times within the hatched region, the current takes relatively similar values. Looking at the yellow circels in the lower panel that represent the weakly bound situation (b) shows that after the (less pronounced) initial rise, also in (b) the first point in time when the initial slope starts to decrease is observed some time before 0.1 fs and within the same hatched region. At approximately the center of the hatched interval the current of the covalently bound configuration (c) reaches its maximum with $I_{\text{lig},\text{c}} \approx  \SI{2.53}{\milli\ampere}$. The weakly bound situation (b) reaches a current of $I_{\text{lig},\text{b}} \approx \SI{0.52}{\milli\ampere}$ before the slope shows first signs of leveling off after the middle of the hatched interval. Thus, there is a factor of ca.\ five difference between the two characteristic currents. 
Following Equation (\ref{eq:SigmaA}), this ratio directly translates into the metric $\sigma A$, which we indicated with a second vertical axis on the right-hand side of the bottom panel of Figure~\ref{fig:IandSigma_E-Field_0.15}. 

For the sake of being able to compare results from the electrical field approach here to the results from the other two approaches that we discuss in the following sections \ref{sec:ConductivityAbsBound} and \ref{sec:condhole}, we define the ratio
\begin{align} \label{eq:SigmaRatio_E-Field}
	\frac{(\sigma \cdot A) |_i }{(\sigma \cdot A)|_\text{c} } \approx	\frac{\sigma_i}{\sigma_\text{c}}
\end{align}
where $i \in \{a, b, c\}$. In this ratio, the unkown cross-sectional area $A$ cancels if one assumes that $A$ is similar in the different configurations. This eases the comparison to the results of the following sections. The time evolution of Equation (\ref{eq:SigmaRatio_E-Field}) is shown in the top panel of Figure~\ref{fig:IandSigma_E-Field_0.15}. Within the hatched area which approximately indicates the range of time in which the ratio is relevant, the ratio takes values around 0.2. This reflects the difference by a factor of ca.\ 5 that was mentioned above.

One might find it dissatisfying that the discussion so far did not give a more precise argument for determining the point in time at which the observed current most closely reflects the connectivity of the ligand-to-cluster anchoring. If desired, one can determine this point in time by making use of the fact that Equations~(\ref{eq:j}) and (\ref{eq:jSigmaE_E-Field}) are based on the linearity of the relation between the external field and the current. 
Thus, one can run further calculations with different values of $E_\mathrm{ext}$.
Doubling the value of $E_\mathrm{ext}$ should double the value of $I_\mathrm{lig}$, as long as the electron dynamics is not governed by other forces than $E_\mathrm{ext}$. The point in time at which the counter-field from the internal polarization becomes relevant can thus be determined as the time at which the linearity between the external field and the current is lost. The largest time at which the linearity still holds is the one at which the connectivity should be evaluated. 
We have found that for times  $ \lesssim \SI{0.1}{\femto\second}$, doubling $E_\text{ext}$ results in a doubling of the current, i.e., the proportionality is observed. This shows that the linearity assumption underlying Equations~(\ref{eq:j}) is correct and that times $ \lesssim \SI{0.1}{\femto\second}$ are appropriate for determing the characteristic $\sigma A$.

\subsection{Analysis of the anchoring with the absorbing boundary method} \label{sec:ConductivityAbsBound}

Next we extend the analysis to the 
case of the absorbing boundary. 
Here, the continuous removal of density from the bottom of the system drives the current by inducing a polarization $\boldsymbol{P}$ that corresponds to a dynamically evolving displacement field  $\boldsymbol{D}$, 
\begin{align} \label{eq:D-Field_Maxwell}
	|\boldsymbol{D}| =   |\boldsymbol{P}|.
\end{align}
The internal
polarization is the 
driving force for the observed flow of charge. 
The polarization builds up because electron density is removed in the bottom region of the system. This removed density corresponds to an amount of charge that can readily be calculated in the real-time simulation by integrating the density over the corresponding volume,
\begin{align}\label{eq:AccumulatedCharges_pos}
	Q_\text{bottom} = |\Delta N_\text{WO$_3$ bulk}| e.
\end{align}
At the top of the system, the outflow of electron density leads to a region that is effectively positively charged, and the charge can again by readily obtained by integrating the density,
\begin{align}\label{eq:AccumulatedCharges_neg}
	Q_\text{top} = |\Delta N_\text{lig}| e.
\end{align}
The net charge difference is
\begin{align}\label{eq:ChargeDifference_AbsBound}
	\Delta Q = Q_\text{bottom} - Q_\text{top} = \left( |\Delta N_\text{WO$_3$ bulk}| - |\Delta N_\text{lig}| \right)  e.
\end{align}
This difference is responsible for the polarization, i.e., a dipole moment per volume
\begin{align} \label{eq:PolarizationDipoles}
	|\boldsymbol{P}| = \frac{\Delta Q \,  d }{2 V}.
\end{align}
Here, $d$ denotes the effective distance of the charge separation that defines the overall dipole that builds up through the action of the absorbing boundary, and $V$ refers to the volume of the system. Both $d$ and $V$ are introduced as auxiliary quantities that will cancel in the final equations, see below. Therefore, one does not need to worry that it might be cumbersome to precisely define such quantities on the atomistic scale.
Since the bottom of the cluster is homogeneously charged and aligned with the $xy$-plane, the polarization is directed along the $z$-axis, and we can reduce all equations to the $z$ component as in the previous case.

The electrical field $E$ is related to the polarization by $E=P/(\epsilon_0 \epsilon)$, where $\epsilon$ would be the permittivity in a macroscopic material. We keep $\epsilon$ in the equations as a material parameter, and as our interest is in determining ratios of the connectivity, the precise definition and value on the atomistic scale is not of importance here. Exploiting 
\begin{equation}
j=\sigma E =\frac{I_\mathrm{lig}}{A} \Rightarrow I_\mathrm{lig}=\sigma A E,
\end{equation}
substituting $P$ for $E$, and using Equation (\ref{eq:PolarizationDipoles}), we arrive at 
\begin{equation}
I_\mathrm{lig}=\sigma A E = \sigma A \frac{\Delta Q d}{2 \epsilon_0 \epsilon V}=
\sigma \frac{\Delta Q}{2 \epsilon_0 \epsilon}.
\end{equation}
In the last step, $V=A d$ was used, and the geometric quantities have thus canceled. As the result we now see that when the dynamics is driven by the absorbing boundary, the quantity
\begin{equation}
\frac{\sigma}{\epsilon_0 \epsilon}=\frac{2 I_\mathrm{lig}}{\Delta Q} 
\end{equation}
can be used as a metric for quantifying the electronic connectivity. The expression on the right-hand side can readily be obtained from the real-time DFT simulation.

By making the assumption that $\epsilon$ will be similar in the different cluster-ligand bonding situations and taking the ratio
\begin{align}\label{eq:SigmaRatio_AbsBound}
	\frac{(I^\text{lig} / \Delta Q)_i}{(I^\text{lig} / \Delta Q)_c} \approx \frac{\sigma_i}{\sigma_c},
\end{align}
we can compare the relative connectivities that we obtain from the absorbing boundary approach to the ones from the electrical field approach.

\begin{figure}
	\centering
	\includegraphics[width=0.7\linewidth]{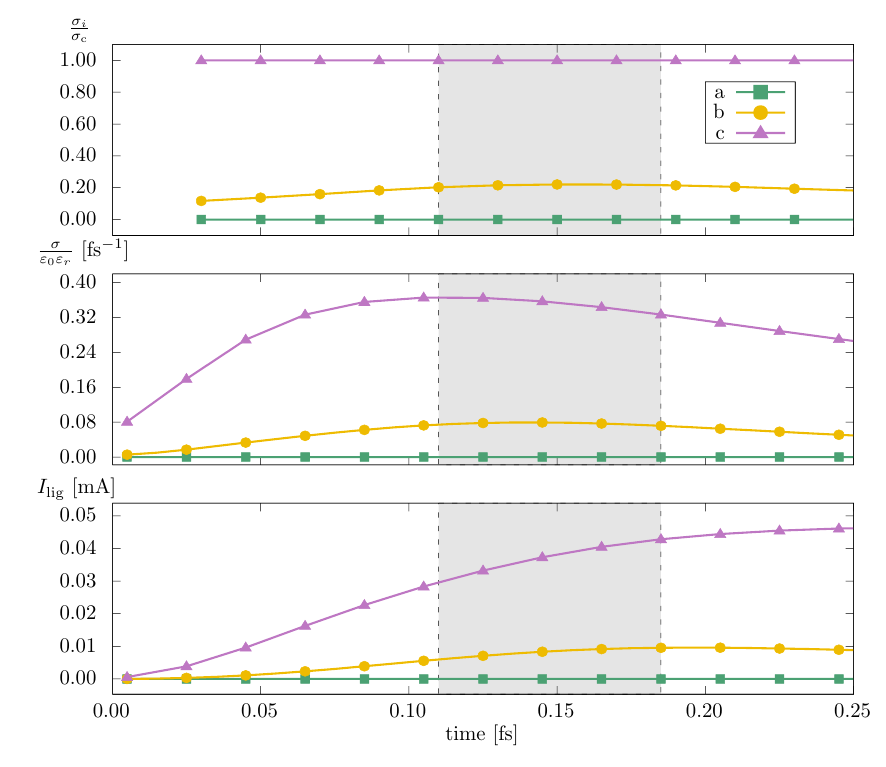}
	\caption{Time evolution of $I_{\text{lig}}$, of  $\sigma / (\varepsilon_0 \varepsilon_r)$, and of their ratio under an absorbing boundary condition. The labels and color codes are the same as in Figure \ref{fig:instIon}. See main text for discussion.
	}
	\label{fig:IandSigma_AbsBound_-4e-}
\end{figure}
\clearpage
\textbf{Figure~\ref{fig:IandSigma_AbsBound_-4e-}} shows $I_{\text{lig}}$ and $\sigma / (\varepsilon_0 \varepsilon_r)$ as a function of time. The color codes and labels are the same as before.
First focusing on the bottom panel we see that the continuous driving of the electron dynamics that the absorbing boundary provides in the configurations (b) and (c) leads to a time-dependence of the current that differs significantly from the one seen with the constant electrical field in the lower panel of Figure~\ref{fig:IandSigma_E-Field_0.15}. This is plausible and expected, because the absorbing boundary prevents the rapid buildup of the counter-field. The data obtained in the absorbing boundary approach are therefore also somewhat easier to interprete. In the far-distant configuration (a), no current is observed, i.e., the sanity check is again fulfilled. The hatched area indicates an interval in which initial effects have died out and the currents have not yet saturated.

The important and decisive information is obtained from the middle and top panel of Figure~\ref{fig:IandSigma_AbsBound_-4e-}. It shows the absolute (middle) and relative (top) connectivity metrics. We see that although the real-time dynamics is beeing initiated in a completely different way than in Section \ref{sec:ConductivityE-Field}, the configuration (c) with two covalent bonds again leads to a ca. five times higher connectivity, clearly reflected by the value of 0.2 that the yellow curve reaches in the hatched interval in the upper panel. This confirms that the obtained ratio is independent of the details of the simulation setups, i.e., the real-time approach indeed allows to quantify the relative electronic conductivity of ligand anchoring.

As a further check we compared different strengths of the absorbing boundary; specifically, one in which 4 electrons were removed from the cluster and another in which 8 electrons were removed. We observed that the current doubles. This linear scaling confirms that the system operates in the regime in which the current is proportional to the number of electrons extracted. 
Consequently, the proportionality factor $\sigma / (\varepsilon_0 \varepsilon_r)$ is independent of the strength of the absorbing boundary. This further validates the robustness and consistency of the approach.

\subsection{Analysis of the anchoring with the hole initialization method}
\label{sec:condhole}

In a similar way as above we can also evaluate the dynamics that follows the initialization of a hole density. The calculation of the conductivity measure is largely analogous to the absorbing boundary approach, with one key distinction: The net charge difference $\Delta Q$ between the top and bottom of the complex is defined differently. After initial electron removal, the system corresponds to a closed system that obeys global charge conservation.

Consequently, the amount of density removed from the ligand, $|\Delta N_\text{lig}|$, is equal to the charge refluxing into the bottom of the cluster, $|\Delta N_\text{WO$_3$ bulk}|$. The net charge difference between the cluster base and the ligand thus corresponds to the positive charge at the base -- arising from the removal of one electron -- corrected by the negative charge that flows back from the top, i.e.,
\begin{align}\label{eq:ChargeDifference_InstIon}
	\Delta Q = \left(1 - |\Delta N_\text{WO$_3$ bulk}| \right)  e,
\end{align}
where $\Delta N_\text{WO$_3$ bulk}$ denotes the change in electron count in the cluster bulk due to the reflux. This charge separation arises almost instantaneously upon the hole initialization and provides a direct measure of the system’s response to the sudden perturbation. From then on, the analysis proceeds analogously to the absorbing boundary case, with the same definitions for the metric of the anchoring conductivity.

\begin{figure}
 	\centering
 	\includegraphics[width=0.7\linewidth]{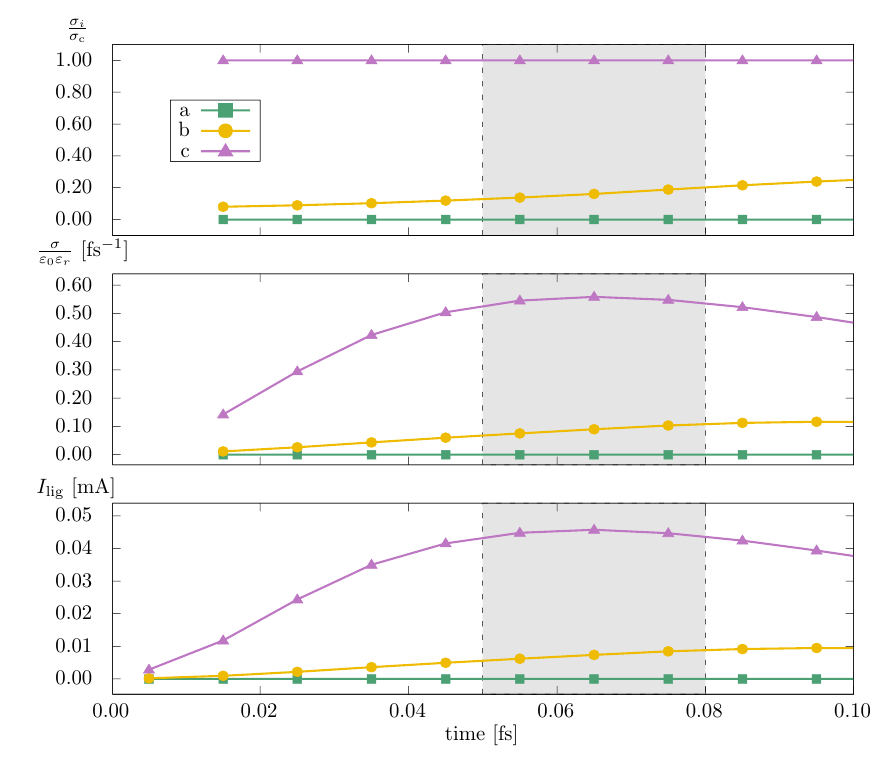}
 	\caption{
 	Time evolution of $I_{\text{lig}}$, of  $\sigma / (\varepsilon_0 \varepsilon_r)$, and of their ratio after hole initialization. The labels and color codes are the same as in Figure \ref{fig:instIon}. See main text for discussion.
 	}
 	\label{fig:IandSigma_instIon}
 \end{figure}
\textbf{Figure~\ref{fig:IandSigma_instIon}} shows the time evolution of the current $I_\text{lig}$, of $\sigma / (\varepsilon_0 \varepsilon_r)$, and of the corrsponding ratios when the ligand-cluster system is subjected to the initialization of a hole density. 
The closed nature of the system necessitates a focus on the first approximately $\SI{0.1}{\femto\second}$ after electron removal to assesss the conductivity measure. Beyond this time, oscillations arising from counter-field effects and interferences distort the response.
Only within the initial ca.\ $\SI{0.1}{\femto\second}$ does the current leaving the ligand plateau to nearly constant values, resulting in smooth conductivity curves and their corresponding ratios.
The hierarchy in the anchoring conductivity that emerges is once more consistent with the previously presented results: 
The current in the configuration (c) is ca.\ five times higher than in configuration (b), and the relative anchoring conductivity is again at the corresponding value of ca.\ 0.2.
As expected, configuration (a) shows no measurable conductivity.

\section{Summary and conclusion}
\label{sec:concl}

In this paper we have introduced three different approaches based on DFT in real-time that allow for characterizing the electronic connectivity between a ligand and the underlying substrate. On the conceptual side, the real-time approach has the big advantage that it relies on evaluating solely the density, i.e., the quantity that becomes exact in DFT for the exact exchange-correlation approximation. It thus completely avoids to endow auxiliary quantities such as eigenvalues or orbital overlap integrals with physical meaning. On the practical side, it has the advantage that the computational efficiency of real-time propagation, e.g., the linear scaling with the number of electrons, can be brought to bear, and one only needs to propagate for short times. A further advantage is that the procedures for quantifying the anchoring quality are straightforward: They only require the evaluation of integrals of the density in certain regions of space. While our work used the real-time propagation in combination with a real-space grid, the methods that we discussed are completely general and can as well be used with other basis sets and implementations of real-time DFT.

We have demonstrated that different anchoring modes -- in our case, anchoring via a hydrogen bond on the one hand, and anchoring via covalent bonds on the other hand -- can be compared quantitatively. We consistently found that for the anchoring of PPA on a WO$_3$ cluster that we studied, the covalent anchoring leads to a five times higher electronic connectivity than the anchoring via a hydrogen bond. We have compared three different ways of initializing the electron dynamics. The hole initialization and the application of an external electrical field lead to complex long-time dynamics, and we have discussed how to choose the time at which the conductivity of the anchoring can be determined reliably. Driving the dynamics via an absorbing boundary leads to smoother signals that make the evaluation easier. All three approaches lead to the same ratio for the connectivity of the hydrogen-bonded and the covalently bonded configuration, establishing consistency.

This paper serves mainly as a proof of concept that a quantitative analysis of ligand anchoring is possible and can readily be realized with DFT in real time. There are several close lying and relevant directions for future extensions of this work. Studying the anchoring on extended surfaces, e.g., via slab calculations, would allow for studying anchoring on crystal surfaces. The electrical field approach that we discussed can be adjusted for slab calculations. One can also think about further ways of initializing the electron dynamics, e.g., by simulating an optical excitation, and one can think about extending such calculations to also include effects of nuclear motion, e.g., by combining the quantum mechanical real-time propagation of the orbitals with classical dynamics for the nuclei. In a yet further step, dissipation could be introduced via suitable operators in the context of stochastic time-dependent density functional theory \cite{agosta2008,eetdirk}. While all these extensions seem worthwhile to develop, it is reassuring that already at the present stage, the relative quality of ligand anchoring for the purposes of transferring charge between a ligand and a substrate can reliably be evaluated on a first-principles basis with the real-time approaches that we discussed here. This will allow theory to support experimental material science further in the development of energy converting systems.

\medskip
\textbf{Acknowledgements} \par %delete if not applicable))
We are grateful to M.\ Knodt for fruitful discussions.
We acknowledge support from project B02 in the Collaborative Research Center 1585 of the German Research Foundation (DFG) (SFB MultiTrans, Grant 492723217), and from the Bavarian State Ministry of Science,  Research, and the Arts for the Collaborative Research Network ``Solar Technologies go Hybrid''. We further
acknowledge computational resources provided by
NHR Erlangen under project ELTRANS (b163cb) within DFG project number 440719683, and by the Bayreuth Centre for High Performance Computing within DFG project number 422127126.

\medskip
\textbf{Date availability statement}
Data underlying this article is provided in the Supplementary Material and can be obtained from the authors at reasonable request. 

\medskip

\bibliographystyle{APPA}
\bibliography{bibliography}

@string{ NANO=  "Nanotechnology"}

@article{kresse1993ab,
	title={Ab initio molecular dynamics for liquid metals},
	author={Kresse, Georg and Hafner, J{\"u}rgen},
	journal={Phys. Rev. B},
	volume={47},
	number={1},
	pages={558},
	year={1993},
	publisher={APS}
}

@article{kresse1996efficiency,
	title={Efficiency of ab-initio total energy calculations for metals and semiconductors using a plane-wave basis set},
	author={Kresse, Georg and Furthm{\"u}ller, J{\"u}rgen},
	journal={Comput. Mater. Sci.},
	volume={6},
	number={1},
	pages={15--50},
	year={1996},
	publisher={Elsevier}
}

@article{kresse1996comput,
  title = {Efficient iterative schemes for ab initio total-energy calculations using a plane-wave basis set},
  author = {Kresse, G. and Furthm\"uller, J.},
  journal = {Phys. Rev. B},
  volume = {54},
  issue = {16},
  pages = {11169--11186},
  numpages = {0},
  year = {1996},
  month = {Oct},
  publisher = {American Physical Society},
  doi = {10.1103/PhysRevB.54.11169},
  url = {https://link.aps.org/doi/10.1103/PhysRevB.54.11169}
}

@article{kresse1996software,
	title={Software {VASP}, {V}ienna, 1999; {G}. {K}resse, {J}. {F}urthm{\"u}ller},
	author={Kresse, G and others},
	journal={Phys. Rev. B},
	volume={54},
	number={11},
	pages={169},
	year={1996}
}

@article{kresse1994norm,
	title={Norm-conserving and ultrasoft pseudopotentials for first-row and transition elements},
	author={Kresse, G and Hafner, J},
	journal={J. Phys.: Condens. Matter},
	volume={6},
	number={40},
	pages={8245},
	year={1994},
	publisher={IOP Publishing}
}

@article{kresse1999ultrasoft,
	title={From ultrasoft pseudopotentials to the projector augmented-wave method},
	author={Kresse, Georg and Joubert, Daniel},
	journal={Phys. Rev. B},
	volume={59},
	number={3},
	pages={1758},
	year={1999},
	publisher={APS}
}

@article{TASK,
	title = {Ultranonlocality and accurate band gaps from a meta-generalized gradient approximation},
	author = {Aschebrock, Thilo and K\"ummel, Stephan},
	journal = {Phys. Rev. Res.},
	volume = {1},
	issue = {3},
	pages = {033082},
	numpages = {14},
	year = {2019},
	month = {Nov},
	publisher = {American Physical Society},
	doi = {10.1103/PhysRevResearch.1.033082},
	url = {https://link.aps.org/doi/10.1103/PhysRevResearch.1.033082}
}

@article{troullier1991efficient,
	title={Efficient pseudopotentials for plane-wave calculations},
	author={Troullier, Norman and Martins, Jos{\'e} Lu{\'\i}s},
	journal={Phys. Rev. B},
	volume={43},
	number={3},
	pages={1993},
	year={1991},
	publisher={APS}
}

@article{kleinman1980relativistic,
	title={Relativistic norm-conserving pseudopotential},
	author={Kleinman, Leonard},
	journal={Phys. Rev. B},
	volume={21},
	number={6},
	pages={2630},
	year={1980},
	publisher={APS}
}

@article{bachelet1982relativistic,
	title={Relativistic norm-conserving pseudopotentials},
	author={Bachelet, Giovanni B and Schl{\"u}ter, M},
	journal={Phys. Rev. B},
	volume={25},
	number={4},
	pages={2103},
	year={1982},
	publisher={APS}
}

@article{SchelterKuemmelBTDFT,
	author = {Schelter, Ingo and K\"ummel, Stephan},
	doi = {10.1021/acs.jctc.7b01013},
	eprint = { https://doi.org/10.1021/acs.jctc.7b01013 },
	journal = {J. Chem. Theory Comput.},
	note = {PMID: 29489356},
	number = {4},
	pages = {1910–1927},
	title = {Accurate Evaluation of Real-Time Density Functional Theory Providing Access to Challenging Electron Dynamics},
	url = {https://doi.org/10.1021/acs.jctc.7b01013},
	volume = {14},
	year = {2018}
}

@article{richter2023,
author={Rian Richter and Thilo Aschebrock and Ingo Schelter and K\"ummel, Stephan},
title = {Meta Generalized Gradient Approximations in Time Dependent Generalized Kohn Sham Theory: Importance of the current-density correction},
journal = {J. Chem. Phys.},
volume = {159},
pages = {124117-1 -- 124117-11},
year = {2023}
}

@article{eel2023,
author={Thilo Aschebrock and Timo Lebeda and Moritz Br\"utting and Rian Richter and Ingo Schelter and K\"ummel, Stephan},
title = {Exact exchange-like electric response from a meta-generalized gradient approximation: A semilocal realization of ultranonlocality},
journal = {J. Chem. Phys.},
volume = {159},
pages = {234107-1 -- 234107-13},
year = {2023}
}

@article{lak1,
author={Timo Lebeda and Aschebrock, Thilo and K\"ummel, Stephan},
title = {Balancing the Contributions to the Gradient Expansion: Accurate Binding and Band Gaps with a Nonempirical Meta-GGA},
journal = {Phys. Rev. Lett.},
volume = {133},
pages = {136402},
year = {2024}
}

@article{lak2,
author={Timo Lebeda and K\"ummel, Stephan},
title = {Meta-GGA that describes weak interactions in addition to bond energies and band gaps},
journal = {Phys. Rev. B},
volume = {111},
pages = {155133},
year = {2025}
}

@article{octopus2020,
author = {Tancogne-Dejean,Nicolas  and Oliveira,Micael J. T.  and Andrade,Xavier  and Appel,Heiko  and Borca,Carlos H.  and Le Breton,Guillaume  and Buchholz,Florian  and Castro,Alberto  and Corni,Stefano  and Correa,Alfredo A.  and De Giovannini,Umberto  and Delgado,Alain  and Eich,Florian G.  and Flick,Johannes  and Gil,Gabriel  and Gomez,Adrián  and Helbig,Nicole  and H\"ubener,Hannes  and Jestädt,René  and Jornet-Somoza,Joaquim  and Larsen,Ask H.  and Lebedeva,Irina V.  and L\"uders,Martin  and Marques,Miguel A. L.  and Ohlmann,Sebastian T.  and Pipolo,Silvio  and Rampp,Markus  and Rozzi,Carlo A.  and Strubbe,David A.  and Sato,Shunsuke A.  and Schäfer,Christian  and Theophilou,Iris  and Welden,Alicia  and Rubio,Angel },
title = {Octopus, a computational framework for exploring light-driven phenomena and quantum dynamics in extended and finite systems},
journal = {The Journal of Chemical Physics},
volume = {152},
number = {12},
pages = {124119},
year = {2020}
}

@article{pbe,
author={J. P. Perdew and K. Burke and M. Ernzerhof},
title={Generalized Gradient Approximation Made Simple},
journal={Phys. Rev. Lett. },
volume={77},
pages={3865},
year={1996}
}

@article{runge1984density,
	title={Density-functional theory for time-dependent systems},
	author={Runge, Erich and Gross, Eberhard KU},
	journal={Phys. Rev. Lett.},
	volume={52},
	number={12},
	pages={997},
	year={1984},
	publisher={APS}
}

@article{manolopoulos2002derivation,
	title={Derivation and reflection properties of a transmission-free absorbing potential},
	author={Manolopoulos, David E},
	journal={J. Chem. Phys.},
	volume={117},
	number={21},
	pages={9552--9559},
	year={2002},
	publisher={American Institute of Physics}
}

@article{dauth2016manoBTDFT,
	title={Perpendicular emission, dichroism, and energy dependence in angle-resolved photoemission: the importance of the final state},
	author={Dauth, Matthias and Graus, Martin and Schelter, Ingo and Wie{\ss}ner, Michael and Sch{\"o}ll, Achim and Reinert, Friedrich and K{\"u}mmel, Stephan},
	journal={Phys. Rev. Lett.},
	volume={117},
	number={18},
	pages={183001},
	year={2016},
	publisher={APS}
}

@article{schaffhauser2016absBound,
	title={Using time-dependent density functional theory in real time for calculating electronic transport},
	author={Schaffhauser, Philipp and K{\"u}mmel, Stephan},
	journal={Phys. Rev. B},
	volume={93},
	number={3},
	pages={035115},
	year={2016},
	publisher={APS}
}

@article{knodt2026determining,
	title = {Determining the binding strength of phenolic anchoring groups on hydrated {WO}3 surfaces},
	issn = {1463-9076},
	doi = {10.1039/d6cp00628k},
	journaltitle = {Physical Chemistry Chemical Physics},
	journal = {Phys. Chem. Chem. Phys.},
	author = {Knodt, Matthias and Mayer, Lukas Max and Gareis, Emmi and von der Haar, Frederike and Hörner, Gerald and Leppert, Linn and Weber, Birgit and Marschall, Roland and Kümmel, Stephan},
	date = {2026-06},
	pages={ 17288–17298},
	year = {2026}
}

@article{ben2022first,
	title={First-Principles Modeling of Dye Anchoring on (001) $\gamma$-Monoclinic WO3 Surfaces: The Role of Oxygen Vacancies},
	author={Ben Jannet, Azza and Said, Moncef and Badawi, Michael and Pastore, Mariachiara},
	journal={J. Phys. Chem. C},
	volume={126},
	number={12},
	pages={5424--5434},
	year={2022},
	publisher={ACS Publications}
}

@article{costa2022current,
	title={Current trending and beyond for solar-driven water splitting reaction on WO3 photoanodes},
	author={Costa, Magno B and de Ara{\'u}jo, Mois{\'e}s A and de Lima Tinoco, Marcos V and de Brito, Juliana F and Mascaro, Lucia H},
	journal={Journal of Energy Chemistry},
	volume={73},
	pages={88--113},
	year={2022},
	publisher={Elsevier}
}

@article{gareis2025submitted,
	author  = {Gareis, Emmi and Schelter, Ingo and K{\"u}mmel, Stephan},
	title   = {Pseudopotentials for meta-GGA calculations},
	journal = {Phys. Rev. B},
	volume={114},
	pages={ 165112-1 --  165112-16},	
	year    = {2026}
}

@article{holliman2018,
  author = {Peter J. Holliman and Christopher Kershaw and Arthur Connell and Eurig W. Jones and Robert Hobbs and Rosie Anthony and Leo Furnell and James McGettrick and Dawn Geatches and Sebastian Metz},
  title = {A perspective on using experiment and theory to identify design principles in dye-sensitized solar cells},
  journal = {Science and Technology of Advanced Materials},
  volume = {19},
  number = {1},
  pages = {599--612},
  year = {2018},
  publisher = {Taylor \& Francis},
  doi = {10.1080/14686996.2018.1492858}
  }

@article{ahsaine2018,
	author  = {M. El ouardi and A. El Idrissi and M. Arab and M. Zbair and H. Haspel and
M. Saadi and H. Ait Ahsaine},
	title   = {Review of photoelectrochemical water splitting: From quantitative approaches to effect of sacrificial agents, oxygen vacancies, thermal and magnetic field on (photo)electrolysis},
	journal = {International Journal of Hydrogen Energy},
	volume={51},
	pages={1044 -- 1067},	
	year    = {2024}
}

@article{li2025,
author = {Li, Chenglong and Mujtaba, Saqib and Quan, Jingjing and Xu, Li and Ning, Xingming and Chen, Pei and An, Zhongwei and Chen, Xinbing},
title = {Activation of Semiconductor/Electrocatalyst/Electrolyte Interfaces Through Ligand Engineering for Boosting Photoelectrochemical Water Splitting},
journal = {Advanced Functional Materials},
volume = {35},
number = {30},
pages = {2501262},
doi = {https://doi.org/10.1002/adfm.202501262},
url = {https://advanced.onlinelibrary.wiley.com/doi/abs/10.1002/adfm.202501262},
eprint = {https://advanced.onlinelibrary.wiley.com/doi/pdf/10.1002/adfm.202501262},
year = {2025}
}

@article{li2020,
author = {Li, Xiaosong and Govind, Niranjan and Isborn, Christine and DePrince, A. Eugene and Lopata, Kenneth},
title = {Real-Time Time-Dependent Electronic Structure Theory},
journal = {Chemical Reviews},
volume = {120},
number = {18},
pages = {9951-9993},
year = {2020}   
}

@book{tddftbook2012,
    editor =    {Miguel A.L. Marques and Neepa T. Maitra and Fernando M.S. Nogueira and E.K.U. Gross and Angel Rubio},
    title =     {Fundamentals of Time-Dependent Density Functional Theory },
    series =    { Lecture Notes in Physics (LNP, volume 83)},
    publisher = {Springer-Verlag Berlin},
    address =   {Heidelberg},
    isbn =      {978-3-642-23517-7},
    edition =   {1},
    year =      {2012}
}

@article{maitra2016,
author = {N. Maitra},
title={Perspective: Fundamental aspects of time-dependent density functional theory},
journal = {J. Chem. Phys.},
volume = {144},
pages = {220901},
year = {2016}
}

@article{cangi2023,
  title = {Electrical conductivity of iron in Earth's core from microscopic Ohm's law},
  author = {Ramakrishna, Kushal and Lokamani, Mani and Baczewski, Andrew and Vorberger, Jan and Cangi, Attila},
  journal = {Phys. Rev. B},
  volume = {107},
  issue = {11},
  pages = {115131},
  numpages = {7},
  year = {2023}
}

@article{calvayrac2000,
author={F. Calvayrac and P.-G. Reinhard and E. Suraud and C. A. Ullrich},
title={Nonlinear electron dynamics in metal clusters},
journal={Physics Reports},
volume={337},
pages={493 -- 578},
year={2000}
}

@article{yabana1996,
author={K. Yabana and G. F. Bertsch},
title={Time-dependent local-density approximation in real time},
journal={Phys. Rev. B },
volume={54},
pages={4484},
year={1996}
}

@article{yabana1999,
author={K. Yabana and G. F. Bertsch},
title={Time-Dependent Local-Density Approximation in Real Time: Application to Conjugated Molecules},
journal={International Journal of Quantum Chemistry},
volume={75},
pages={55-66},
year={1999}
}

@article{jakowski2025,
    author = {Jakowski, Jacek and Lu, Wenchang and Briggs, Emil and Lingerfelt, David and Sumpter, Bobby G. and Ganesh, Panchapakesan and Bernholc, Jerzy},
    title = {Simulation of 24,000
Electron Dynamics: Real-Time
Time-Dependent Density Functional Theory (TDDFT) with the Real-Space
Multigrids (RMG)},
    journal = {Journal of Chemical Theory and Computation},
    volume = {21},
    number = {3},
    pages = {1322-1339},
    year = {2025},
    month = {02}
}

@article{zelovich2014,
author = {T. Zelovich  and L. Kronik and O. Hod},
journal = {J. Chem. Theory Comput.},
title={State Representation Approach for Atomistic Time-Dependent Transport Calculations in Molecular Junctions },
volume = {10},
pages = {2927 -- 2941},
year = {2014}
}

@article{oz2023,
    author = {Oz, Annabelle and Nitzan, Abraham and Hod, Oded and Peralta, Juan E.},
    title = {Electron Dynamics
in Open Quantum Systems: The Driven
Liouville-von Neumann Methodology within Time-Dependent Density Functional
Theory},
    journal = {Journal of Chemical Theory and Computation},
    volume = {19},
    number = {21},
    pages = {7496-7504},
    year = {2023}
}

@article{andermatt2018,
    author = {Andermatt, Samuel and Bani-Hashemian, Mohammad Hossein and Ducry, Fabian and Brück, Sascha and Clima, Sergiu and Pourtois, Geoffrey and VandeVondele, Joost and Luisier, Mathieu},
    title = {Microcanonical RT-TDDFT simulations of realistically extended devices},
    journal = {The Journal of Chemical Physics},
    volume = {149},
    number = {12},
    pages = {124701},
    year = {2018}
}

@article{he2018,
    author = {He, Shenglai and Covington, Cody and Varga, Kálmán},
    title = {Study of local currents in low dimension materials using complex injecting potentials},
    journal = {Journal of Applied Physics},
    volume = {123},
    number = {16},
    pages = {165102},
    year = {2018}
}

@article{diventra2004,
author = {M. Di Ventra and T. N. Todorov},
title={Transport in nanoscale systems: the microcanonical versus grand-canonical picture},
journal = {J. Phys.: Condens. Matter},
volume = {16},
pages = {8025 -- 8034},
year = {2004}
}

@article{agosta2008,
author = {R. D'Agosta and M. Di Ventra},
title={Stochastic time-dependent current-density-functional theory: A functional theory of open quantum systems},
journal = {Phys. Rev. B},
volume = {78},
pages = {165105},
year = {2008}
}

@article{cheng2006,
author = {Chiao-Lun Cheng and Jeremy S. Evans and Troy Van-Voorhis},
issn = {1098-0121},
journal = {Phys. Rev. B},
number = {15},
pages = {155112},
title = {{Simulating molecular conductance using real-time density functional theory}},
volume = {74},
year = {2006}
}

@article{evans2009nanolett,
author = {J. S. Evans and T. Van Voorhis},
title = {Dynamic Current Suppression and Gate Voltage Response inMetal-Molecule-Metal Junctions},
journal = {Nano Lett.},
volume = {9},
pages = {2671},
year = {2009}
}

@article{toher2005,
author={C. Toher and A. Filippetti and S. Sanvito and K. Burke},
title={Self-Interaction Errors in Density-Functional Calculations of Electronic Transport},
journal={Phys. Rev. Lett.},
volume={95},
pages={146402-1--146402-4},
year={2005}
}

@article{ke2007,
author = {S.-H. Ke and H. U. Baranger and W. Yang},
title={Role of the exchange-correlation potential in ab initio electron transport calculations},
journal = {J. Chem. Phys.},
volume = {126},
pages = {201102},
year = {2007}
}

@article{sicinteger,
author={Hofmann, D. and K{\"u}mmel, S.},
title={Integer particle preference during charge transfer in Kohn-Sham theory},
journal={Phys. Rev. B},
volume={86},
pages={201109(R)},
year={2012}
}

@article{eetdirk,
author={Hofmann-Mees, D. and H. Appel and M. Di Ventra and K{\"u}mmel, S.},
title={Determining Excitation-Energy Transfer Times and Mechanisms from Stochastic Time-Dependent Density Functional Theory},
journal={J. Phys. Chem. B},
volume={117},
pages={14408},
year={2013}
}

@article{trepl2025time,
	title={Time-Resolved Ultrafast Excitation Dynamics in the {B}850 Light-Harvesting Antenna from Density Functional Theory},
	author={Trepl, Thomas and Schelter, Ingo and K{\"u}mmel, Stephan},
	journal={J. Phys. Chem. Lett.},
	volume={16},
	number={42},
	pages={10891--10898},
	year={2025},
	publisher={ACS Publications}
}

\clearpage

\begin{figure}
\textbf{Table of Contents}\\
\medskip
We present three different methods based on density functional theory in real time that allow for quantifying the efficiency with which charge can be transferred between a ligand and a substrate to which the ligand is anchored. For the case of phenylphosphonic acid attached to a WO$_3$ cluster, the anchoring via a hydrogen bond leads to a five times weaker electronic connectivity than covalent anchoring.  

 \centering
  \includegraphics{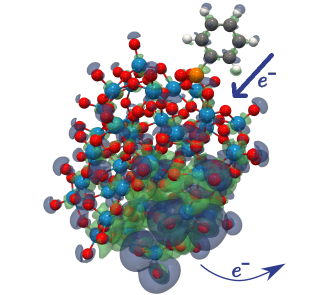}
\end{figure}

\end{document}